\def \SAIT #1 #2 {{\em Mem.\ Soc.\ Astron.\ It.\/} {\bf #1}, #2}
\def \MESS #1 #2 {{\em The Messenger\/} {\bf #1}, #2}
\def \ASTRNACH #1 #2 {{\em Astron. Nach.\/} {\bf #1}, #2}
\def \AAP #1 #2 {{\em Astron. Astrophys.\/} {\bf #1}, #2}
\def \AAL #1 #2 {{\em Astron. Astrophys. Lett.\/} {\bf #1}, L#2}
\def \AAR #1 #2 {{\em Astron. Astrophys. Rev.\/} {\bf #1}, #2}
\def \AAS #1 #2 {{\em Astron. Astrophys. Suppl. Ser.\/} {\bf #1}, #2}
\def \AJ #1 #2 {{\em Astron. J.\/} {\bf #1}, #2}
\def \ANNREV #1 #2 {{\em Ann. Rev. Astron. Astrophys.\/} {\bf #1}, #2}
\def \APJ #1 #2 {{\em Astrophys. J.\/} {\bf #1}, #2}
\def \APJL #1 #2 {{\em Astrophys. J. Lett.\/} {\bf #1}, L#2}
\def \APJS #1 #2 {{\em Astrophys. J. Suppl.\/} {\bf #1}, #2}
\def \APSS #1 #2 {{\em Astrophys. Space Sci.\/} {\bf #1}, #2}
\def \ASR #1 #2 {{\em Adv. Space Res.\/} {\bf #1}, #2}
\def \BAIC #1 #2 {{\em Bull. Astron. Inst. Czechosl.\/} {\bf #1}, #2}
\def \JSQRT #1 #2 {{\em J. Quant. Spectrosc. Radiat. Transfer\/} {\bf #1}, #2}
\def \MN #1 #2 {{\em Mon. Not. R. Astr. Soc.\/} {\bf #1}, #2}
\def \MEM #1 #2 {{\em Mem. R. Astr. Soc.\/} {\bf #1}, #2}
\def \PLR #1 #2 {{\em Phys. Lett. Rev.\/} {\bf #1}, #2}
\def \PASJ #1 #2 {{\em Publ. Astron. Soc. Japan\/} {\bf #1}, #2}
\def \PASP #1 #2 {{\em Publ. Astr. Soc. Pacific\/} {\bf #1}, #2}
\def \NAT #1 #2 {{\em Nature\/} {\bf #1}, #2}

\documentstyle{memsait}
\input epsf.sty
\begin{opening}
\title{PRE-SUPERNOVA RING AROUND PSR0540-69.} 
\author{PATRIZIA A. CARAVEO$^{1,2}$,  ROBERTO MIGNANI$^3$ and GIOVANNI
F. BIGNAMI$^{4,5}$ }
\institute{$^1$Istituto di Fisica Cosmica del CNR , Milano, Italy\\
$^2$Istituto Astronomico, Roma, Italy\\
$^3$STECF-ESO, Garching, Germany \\
$^4$Agenzia Spaziale Italiana, Roma, Italy \\
$^5$Universit\'a di Pavia, Italy
}
\date{} 
\end{opening}

\begin{document}

\oddpagefooter{}{}{} 
\evenpagefooter{}{}{} 
\ 
\bigskip

\begin{abstract}
SNR0540-69 is   a supernova  remnant in the   LMC, harbouring  a young
($\tau \sim 1600~ yrs$) radio/optical/X-ray pulsar (P=50  ms). Ground
based $H\alpha$ imaging of the region has shown a unique spiral-like
structure centered  around   the
pulsar.  In narrow band HST imaging, the feature seems
resolved in a ring-like structure,  probably 
ejected by the progenitor star in a pre-supernova phase 
($\ge 10^{4}~yrs$ ago).

\end{abstract}

\section{Introduction.}
SNR0540-69 is the remnant of the  last supernova explosion occurred in
the Large  Magellanic  Cloud before SN1987A  (Kirshner   et al, 1989).  
High resolution  $H\alpha$ images of the remnant
unveiled a spiral-like  feature, centered on the
pulsar   optical  counterpart, and definitely not present in other wide (B and
V) or narrow-band filter (OIII and SII) images (Caraveo et  al,
1992). 
The   spiral-like feature  was later
confirmed by a deeper $H\alpha$ exposure, 
taken with sub-arcsec seeing conditions. 
The resulting image is shown in Figure 1.\\
To investigate the nature of the $H\alpha$ structure, we have pursued both
spectroscopy and 
high resolution, narrow band imaging.
 
\begin{figure}
\epsfysize=6cm 
\hspace{4.0cm}\epsfbox{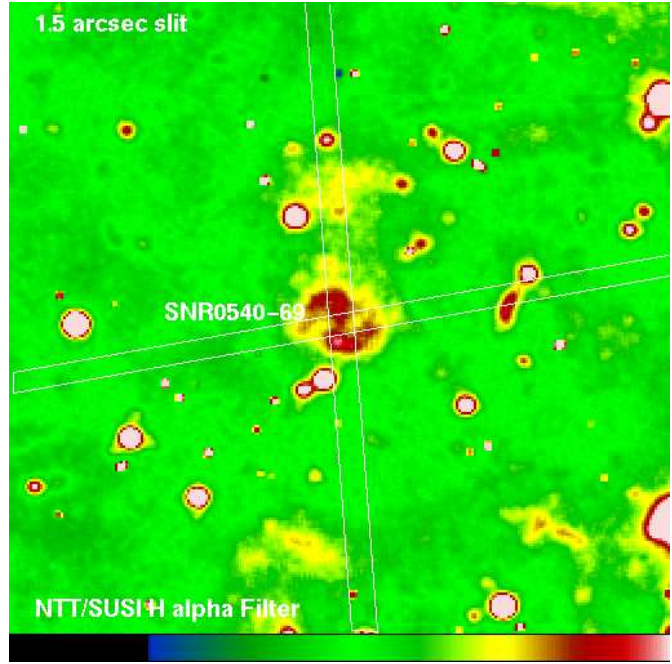}  
\vspace{1.5cm}
\caption[h]{30 min $H\alpha$ exposure of SNR0540-69 taken in Jan 1994 
with the SUperb Seeing Imager (SUSI) of the ESO/NTT
and a $\sim$ 0.6" seeing.  
North to the top, East  to the left.  The spiral-like feature is
$\sim$  4" 
in size. The two stripes
 show the slit orientations.} 
\end{figure}

\section{The data.}

\subsection{Spectroscopy.}

Spectroscopy of SNR0540-69 was performed on 1995 January, at ESO 
(La Silla) using EMMI ({\it ESO Multi Mode
Instrument}) at the 3.5m NTT.  Two
50-minute spectra were  obtained with a  1.5" slit  centered at the  pulsar
position and the long axis oriented North-South and East-West,
as shown in Fig.1. 

\begin{figure}[h]
\epsfysize=5.cm 
\hspace{.0cm}\epsfbox{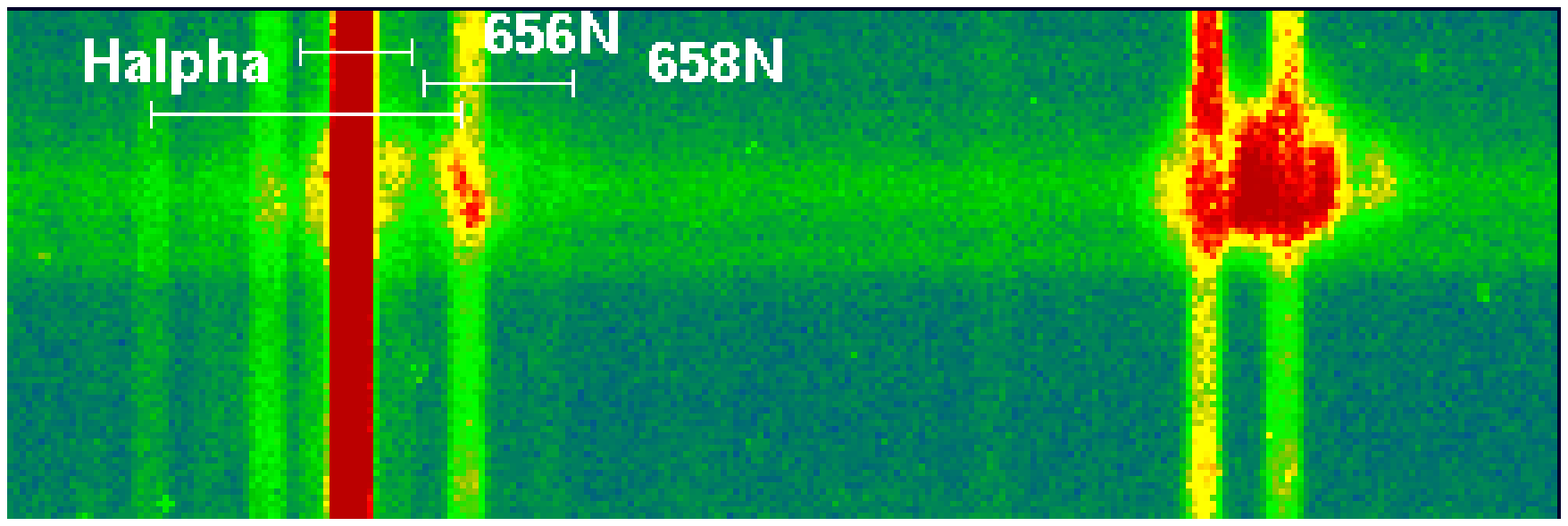 } 
\epsfysize=5.cm 
\hspace{.0cm}\epsfbox{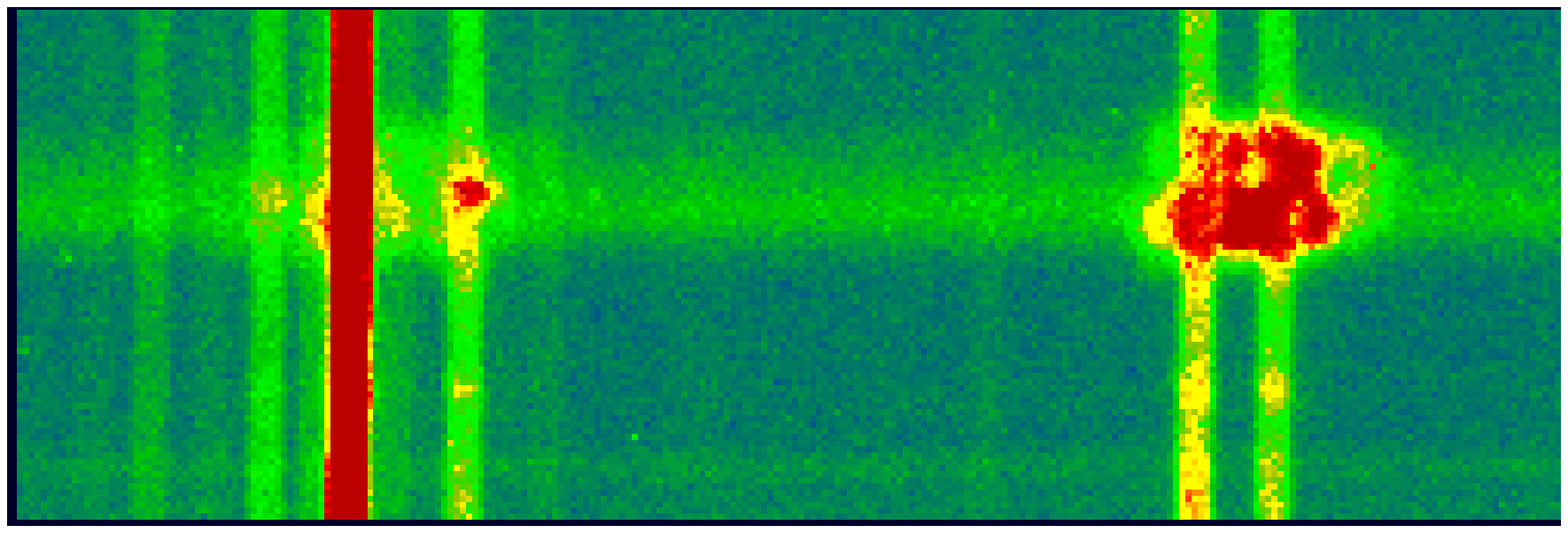} 
\caption[h]{Blow-up of the 2-D  spectra, showing the $H\alpha$ complex
and the SII doublet, for the N-S (a) and E-W (b) orientations of
the slit.
x-axis: wavelengths, increasing from left to right (1 pixel = 1.2 $\AA$);
y-axis: spatial coordinate along the slit (1 pixel=0.27")
	(a) North at top, South at bottom
	(b) East at top, West at bottom.
The passbands of the HST filters 
656N ($\lambda=6562 \AA;\Delta \lambda= 22 \AA$) and
 658N ($\lambda=6590 \AA;  \Delta \lambda= 28.5 \AA$) are also shown
together with the ESO/NTT $H\alpha$ ($\lambda=6552 \AA; \Delta 
\lambda= 60 \AA$). }
\end{figure}

Globally, our results are in excellent agreement with the
comprenhensive work of  Kirshner et al (1989) but for the  region
around $H\alpha$.  The identification of the feature 
observed at $\sim 6580 \AA$ is not straightforward. It was first 
attributed to [NII] ($6584~\AA)$ by
Mathewson et al (1980), then to CaI  ($6572.8~\AA)$ by Dopita \& Tuohy
(1984)  and,  finally, to  $H\alpha$  ($6562.8~\AA)$ by Kirshner et al
(1989).  However, the width of the feature  ($\sim 3500 \pm 50~km/s$ as
 opposed to the values of $\sim
1800-2000~km/s$ measured for single unblended lines) is far too large
to allow the identification with a single line.  
In order to better assess the structure of the of the $6580 \AA$ region
we can exploit the excellent quality of our 2-D spectra.
Figs.2a and b show a blow-up of the NS
and EW 2-D spectra  in the region 6500-7000  $\AA$ including  both the
SII  doublet  and the  $6580  \AA$ region.    While the SII  lines are
blended and significantly redshifted (to  wit their origin in a  young
SNR which is both expanding at $v \sim 1200~km/s$ and receding at 
$v \sim 600~km/s$), the $6580 \AA$ complex is
composed of three lines which appear both narrower and less (if at all)
redshifted.  As in  the Crab Nebula (e.g.  Nasuti  et al.  1996),  we
identify the lines as an $H\alpha$,  unrecoverably  polluted  by LMC
emission, in the  middle of a better  defined NII doublet ($6548, 6584
\AA$). The three lines appear unblended and the troughs, although
partially filled by the synchrotron  continuum  of the  plerion, are
deeper than expected if the lines were affected by the same velocity
dispersion as the SII.  Moreover, looking at the 
brighter $6584 \AA$ component of the NII doublet, we see that, 
while  its  appearence is markedly different in the two spectra, its
centroid is not significantly redshifted w.r.t. the LMC lines. 
However, the structure of the  $6584 \AA$ line points
towards a non-spherical emitting region which is not expanding at the
rate measured for SNR0540-69  neither is it receding  at the  same
speed. 

\subsection{Imaging.}

New imaging of SNR0540-69 was performed in October 1995 with the
HST/WFPC2. To image the different components of the $\lambda=6580 \AA$
feature, clearly  blended  in the wider ESO $H\alpha$
filter (see the passband schematically shown in Fig. 2a), we have used
the WFPC2  "$H\alpha$"   (656N) and  ``N[II]''   (658N)
filters.  The  resulting
images are shown as contour plots in Fig.3 a,b where all field  stars,
falling  within or  close   the remnant,  were removed.  
The difference in the
remnant brightness between  the two filters confirms our spectroscopy,
where the NII [$6584  \AA$]  line  appeared most  prominent  than the
$H\alpha$  one.   The most notable   feature in Fig.  3a,b is a
bright emission knot $\sim  1''$ south  of the  pulsar
position, clearly visible also in the $H\alpha$ NTT image
(Fig.1). Since the  knot does  not appear in our reference HST V-band
image,  we attribute it to the remnant structure  rather
than to a fore/background  object.   Although the S/N  is too  low  to
claim any definite association, a physical  link between this knot and
PSR0540-69 seems to be present  in the  656N  image  (Fig.3a).  The
interaction of PSR0540-69 with the surrounding medium is clearly shown
in the 658N image  (Fig.3b) where a conical structure
is  seen  to  originate from   the pulsar and  to   connect it with an
arc-shaped feature   $\sim 1"$   to the NW.  A  similar, albeit
fainter,  arc is visible  on   the East side of    the pulsar and  the
combination of the two  yields a ring-like  structure ($\sim$ 3"
in  diameter and  $\sim$   0.2"   thick)  centered  roughly  on
PSR0540-69.  An additional,  faint North-South, bar-like,  emission is
seen   to  go  through,  or originate   from,    the pulsar.  Although
definitely brighter in the  "[NII]  filter", a ring-like structure  of
similar dimensions can be inferred in the noisy 656N image (Fig. 3a).

\begin{figure}
\epsfysize=6cm 
\hspace{0.1cm}\epsfbox{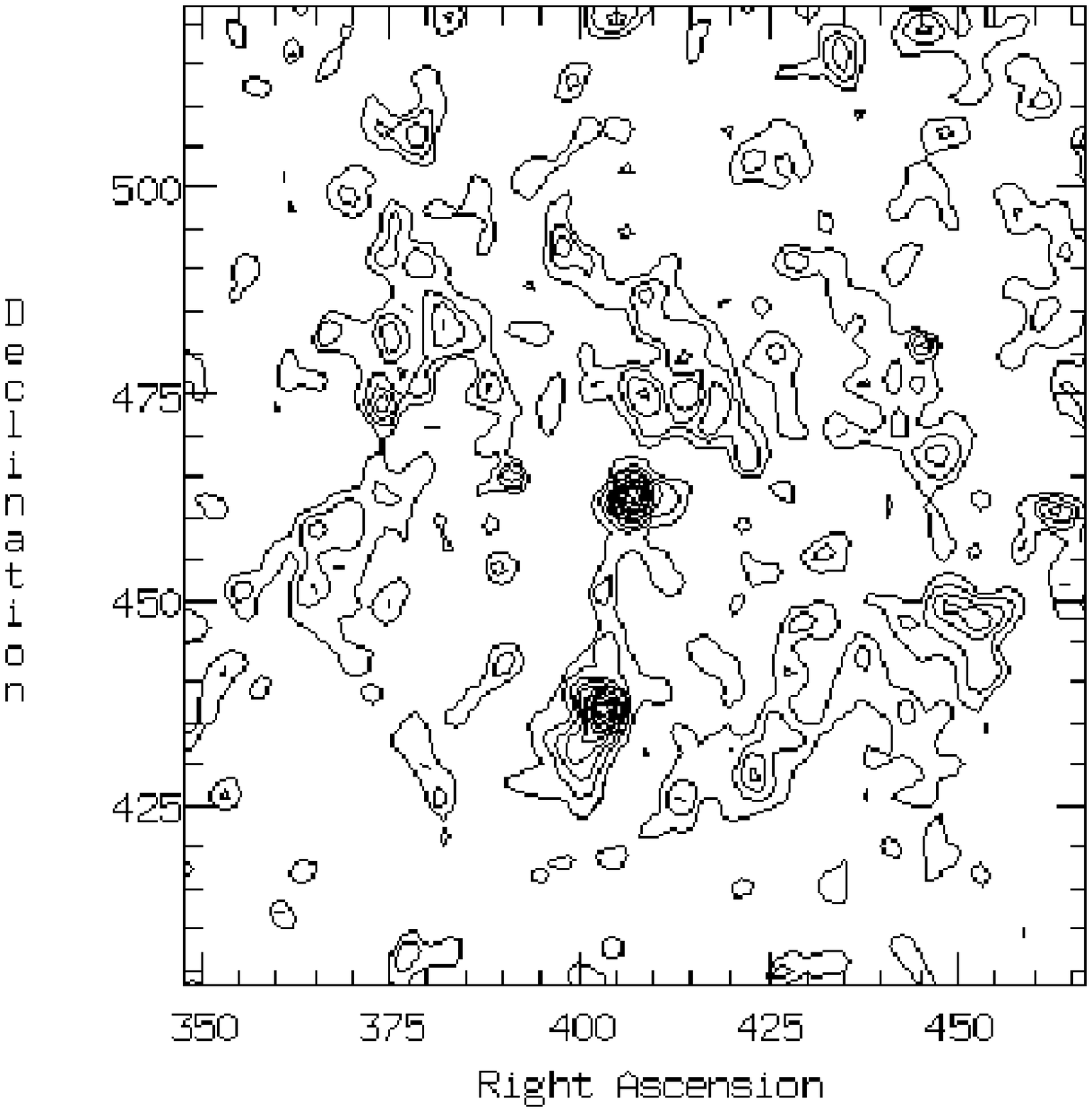} 
\epsfysize=6cm 
\hspace{0.1cm}\epsfbox{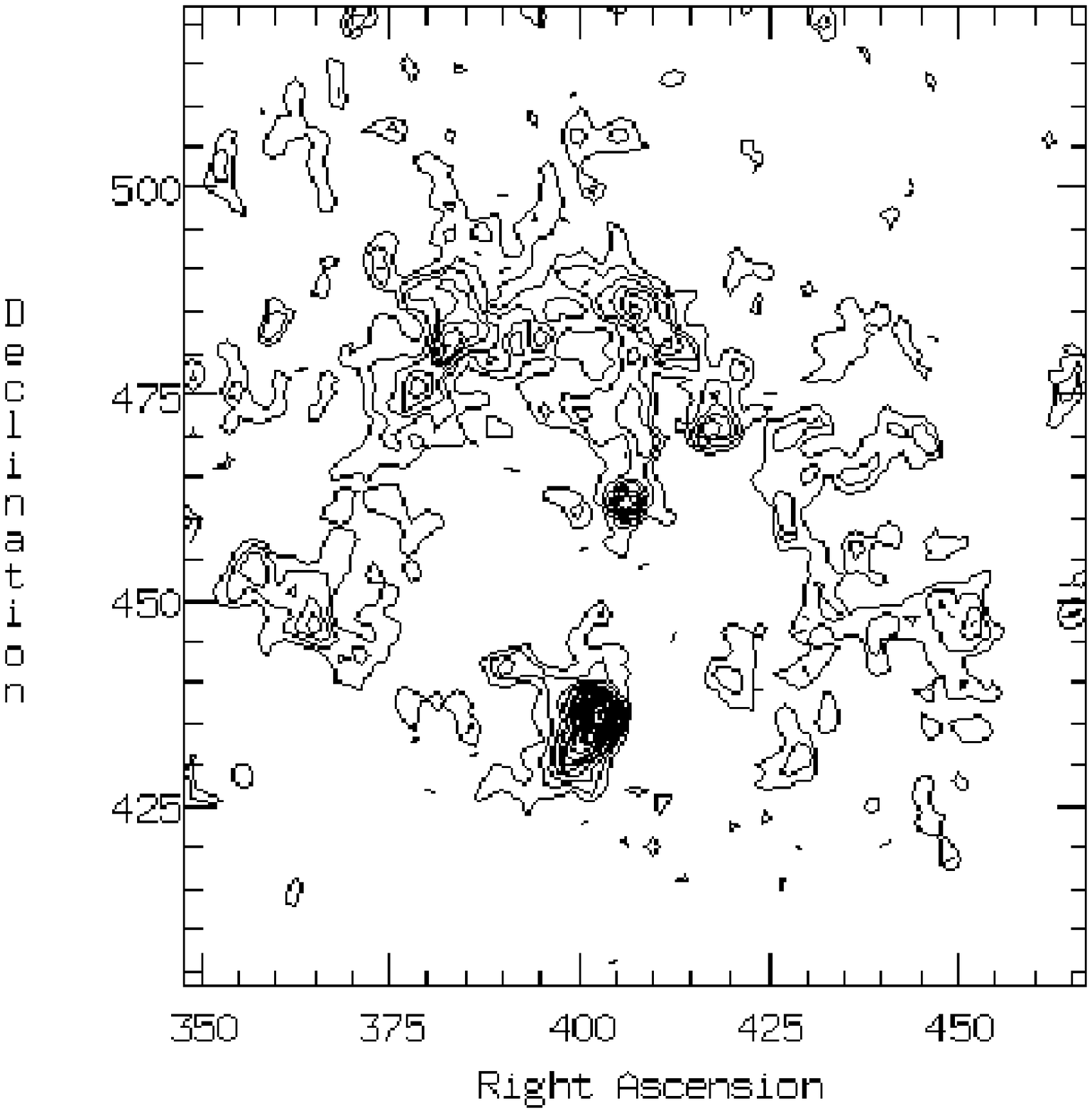}
\caption[h]{Contour plots of WFPC2 images of SNR0540-69:  
(a) 656N "$H\alpha$ filter" ($\lambda=6562 \AA;
\Delta \lambda= 22 \AA$), t=3400s; 
(b) 658N "$N[II]$ filter" ($\lambda=6590 \AA; \Delta
\lambda= 28.5 \AA$),t=4000s. 
The frames are aligned in RA and DEC.  Axis units
are PC pixels (0.045 ") for a corresponding field of view  of
about  5.5'' $\times$5.5".  Each isophote corresponds 
to an increment  of  0.2 counts/pixel against a background level of 1. Point 
sources falling within or near the SNR have been removed via PSF subtraction.}
\end{figure}

\section{Discussion.}

Since the passbands of   the   656N  and  658N  filters  encompass
respectively the $H\alpha$  and  [NII] line,  we can directly compare the 
spectra, shown in Fig.2 a,b, with the HST images, shown in Fig. 3 a,b, once
the different pixel sizes of the two intrument are properly accounted for. 
Since the NTT/EMMI spatial resolution (y-axis on Fig 2) is 0.27'' per pixel
and the PC  pixel size is 0.045'', the 1.5"
slit covers $\sim  33$ PC  pixels.    
In the N-S  spectrum, shown in Fig.2a, the [NII]
line has  a spatial extent identical to
the ring diameter and its brightest part corresponds to the southern knot. 
Going from  North  to South, the average  velocity  of the line shifts
redward continously  for  a  total    displacement of  3-4 EMMI  pixels,
corresponding to $\le 5 \AA$.  In the E-W  spectrum, the [NII] line is
dominated by the emission of the knot, now appearing in the upper part
of Fig.2b, owing to the rotation of the CCD.  The less intense (and
slightly  bluer)  part of the   line  is due to   the SW  part  of the
arc-shaped  feature, which we have   tentatively interpreted as a ring
around  the pulsar.\\ All  in all, the velocity  of  the [NII] feature
differs very little from  the LMC one and  the overall shear along the
N-S  diameter can be  quantified to be  $\le 100~ km/s$.  However, a
ring expanding $\le 50 ~km/s$ could not have been formed at the time
of the SN explosion. To account for  the ring dimensions, an expansion
velocity of $\sim 200 ~km/s$ would be  needed and our data do not
support a velocity shear of $\sim 20 \AA$, corresponding to 16 EMMI pixels. 
Moreover, the [NII] line emission originated from the ring is not
affected by the overall SNR0540-69 recession speed. 
Therefore this "ring" must have formed before the event responsible
for such a recession, i.e. before the SN event itself. \\
Our slowly expanding  ring would thus trace  a pre-supernova  stage of
the progenitor star like, probably,  for the rings
around SN1987A (e.g.  Plait et al,1995, Panagia et al, 1996) and 
for the hourglass nebula around the blue supergiant (BSG) Sher 25 
(Brandner et al, 1997a and b).  SNR0540-69 would thus be
the  second example, after  SN1987A,  of a supernova remnant retaining
memory of a pre-supernova  mass ejection similar  to that undergone by
the BSG Sher25 $\sim$ 6,600 year ago (Brandner et al, 1997b).  Indeed,
both the dimensions and the slow expansion rate of the SNR0540-69 ring
are reminiscent of the inner ring of SN1987A as  well as of that around
Sher 25.  \\
 If we now
concentrate on the comparison between SNR0540-69 and SN1987A, we find 
that  their many similarities  in dimensions, composition and
velocity point towards  similar   progenitors for these   two supernovae
which happen to  be in   the same  star  forming  region in the    LMC
(Kirshner et  al,  1989).  Assuming that  the  progenitor of the $\sim
1600~yrs$  old  SNR0540-69 experienced   mass ejection $\ge 10^{4}~yrs$
before the supernova explosion and  using  the distance value of  50.9
kpc, as inferred  by Panagia et   al (1997) for  SN1987A, the measured
angular  size of the ring ($\sim$  3")  implies a velocity $\sim
30-40~km/s$. This is   larger to the expansion  velocity measured
for the inner ring  of SN1987A ($\sim  11 ~km/s$, Panagia et  al,
1996) but similar  to  the value of  $\sim  30 km/s$ found for   the
$\sim$ 6,600 y old ring around Sher25 (Brandner et al, 1997b). 

\section{Conclusions}

Coupling high resolution imaging with spectroscopy, we have shown that
the  ring-like structure around   PSR0540-69  was formed during   the
pre-supernova phase and survived to the present  day.  We can exclude
that  the ring be  simply due to SN  ejecta since this would imply an
expansion velocity  too high to  be consistent with our  spectral data
but  far too low to  be   consistent with  the average expansion  rate
measured   for the SNR.   Moreover,  the  absence  of any  significant
redshift for the NII line argues strongly  in favor of a pre-supernova
origin.  In the  assumption that the  receding speed of SNR0540-69 was
acquired during  the  supernova   explosion,  the "ring"  must   have
originated before such  explosion,  during the pre supernova  phase of
the progenitor as it is believed  to be the case  for the rings around
SN1987A. \\
However, considering also the case of the ring detected around the BSG
Sher 25, we  can study such structures well  before  the SN explosion,
and follow them when they  are lighted first by  the SN flash and than
by the interaction with the  SN ejecta and,  possibly, with the pulsar
relativistic  wind. Indeed,  at  variance with the  case of SN1987A,
where, as  expected  (Luo et  al 1994), the   inner ring is  now being
reached   by the   supernova ejecta  (NASA   PR  98-24), the ring   of
SNR0540-69 has been swept over by the ejecta long ago and it is now 
shining after $\sim 1,600~yrs$ from the SN explosion most probably owing
to   shock heating. On  the other hand,  the
presence of PSR0540-69, i.e. a young energetic ($\dot E \sim 1.5~10^{38}$
erg/s) pulsar powering the surrounding plerion, could also play an active
role.


\end{document}